\documentclass[aps,prx,twocolumn,superscriptaddress]{revtex4-2}
\usepackage[utf8]{inputenc}
\usepackage{xcolor}
\usepackage{graphicx}
\usepackage{float}
\usepackage{latexsym}
\usepackage{amsmath}
\usepackage{amsfonts}
\usepackage{amssymb}
\usepackage{bm}
\usepackage{calc}
\usepackage{mathtools}
\usepackage{amsbsy}
\usepackage{physics}
\usepackage{enumitem}
\usepackage{media9}
\usepackage{hyperref}
\graphicspath{\figures}

\newcommand{\cblue }{\color{black}}
\newcommand{\cmag }{\color{black}}
\newcommand{\lla}{\left\langle}
\newcommand{\rra}{\right\rangle}

\begin{document}

\title{Self-Avoidance Governs Re-entrant Coil–Globule Transition of Inertial Active Brownian Ring Polymers }

\author{Rakesh Palariya}%
\affiliation{Department of Physics, Indian Institute of Science Education and Research, \\ Bhopal 462 066, Madhya Pradesh, India}

\author{Arindam Panda}%
\affiliation{Department of Physics, Indian Institute of Science Education and Research, \\ Bhopal 462 066, Madhya Pradesh, India}

\author{Roland G. Winkler}%
\affiliation{Theoretical Physics of Living Matter, Institute for Advanced Simulation, Forschungszentrum J{\"u}lich, 52425 J{\"u}lich, Germany}

\author{Sunil P Singh}
 \email{spsingh@iiserb.ac.in}
\affiliation{Department of Physics, Indian Institute of Science Education and Research, \\ Bhopal 462 066, Madhya Pradesh, India}

\begin{abstract}
Active matter systems composed of self-propelled entities exhibit a wide range of out-of-equilibrium phase behaviors and collective dynamics. Assemblies of active particles, e.g., in a linear structure, are particularly interesting, as they exhibit an intriguing interplay between activity and geometrical restrictions due to self-avoidance and entanglements.
Using Langevin dynamics simulations of an active Brownian ring polymer with translational and rotational inertia, we uncover an inertia-driven re-entrant coil–globule transition in the absence of any attractive interactions. We show that rotational inertia plays a paramount role, leading, in the globular state, to spontaneous, nonzero, radially outward polarization of the active forces. {\cblue Self-avoidance and entanglements break the rotational symmetry of active beads and effectively stabilize the globular states.} The transition becomes increasingly sharp with increasing rotational inertia, yielding highly compact globular conformations at intermediate activities, whereas at high activity the polymer re-expands into a coil state. Scaling of the radius of gyration reveals crossovers from self-avoiding, phantom-like to compact-globular structures. Further, rotational inertia enhances transport properties, with an intermediate activity-induced Rouse-type bead diffusion and a long-time diffusion coefficient displaying a power-law dependence on the moment of inertia. Our study establishes a purely inertia-induced mechanism for controlling the collapse of active polymers, arising from spontaneous active-force polarization and self-avoidance.

\end{abstract}
\maketitle 

\section{Introduction}
The coil–globule transition of polymers is pivotal in a broad range of applications, including stimuli-responsive polymer systems, smart materials, drug delivery, and bio-signaling in transcription and gene regulation~\cite{mukherji2014polymer,ulianov2016active,goychuk2024delayed}. Polymer conformational responses to external control parameters—such as electrostatic interactions~\cite{barrat1996theory,wink:98,dobrynin2005theory,muthukumar201750th}, temperature \cite{doi:86,rubi:03,muthukumar2016polymer}, and co-solvency~\cite {mukherji2014polymer,jia2016re,macc:16} have been extensively addressed. In recent years, theoretical and simulation studies have revealed that nonthermal, active processes, either internal or by the embedding environment, profoundly influence the conformational and dynamical properties of polymers and filaments  \cite{wink:20,chelakkot2014flagellar,ghosh2014dynamics,wang2015one,kaiser2015does,eise:16,bianco2018globulelike,anan:18,prathyusha2018dynamically,sain:18.1,martin2019active,mous:19,anan:20,liao2020extensions,mousavi2021localization,philipps2022tangentially,kumar2023local}. In particular, the nonequilibrium fluctuations can cause polymer shrinkage or swelling\cite{anan:18,loca:21,anan:20,pand:25.1,kumar2023local}. 

Active processes play a crucial role in biological systems, from the micro to the macro level. A particular example is the chromosome organization and chromatin folding/unfolding in the cell interior, processes that regulate transcription~\cite{goychuk2023folding,chan2024activity,ulianov2016active} and are crucial for maintaining a gene's function~\cite{hager2009transcription}. The packing of DNA in the cell nucleus is a complex, hierarchical process, from the formation of nucleosome-core particles up to the folding into chromosomes~\cite{albe:13,tubi:24}. Different mechanisms lead to the packing of the circular DNA of prokaryotes into a nucleoid via supercoiling~\cite{song:15,tubi:24}. These processes are guided by proteins that bind to DNA molecules.  In the strive to develop a synthetic cell, which exhibits features of biological ones~\cite{guin:22,giav:25}, compaction of macromolecules may also be required.
Here, other non-protein-mediated active mechanisms may be exploited based on the properties revealed by the above-mentioned theoretical and simulation studies. 

{\cblue Studies of filamentous-type systems have focused on the overdamped dynamics. Yet, subsequent investigations incorporating inertial effects have revealed a profound impact of inertia on the conformational and dynamical properties of active polymers~\cite{teje:24,faze:23,buglakov2025motility,karan2024inertia}. Even more, inertia governs collective migration, stability, maneuverability, and coordinated patterns across scales and systems of fish, insects, ants, birds, snakes, worms, and nematodes~\cite{deblais2023worm,mukundarajan2016surface,cavagna2010scale,cavagna2014bird,marvi2014sidewinding,deblais2020rheology,back:13,malv:19}. 
Synthetic and biological microswimmers exhibit inertial responses driven by rapid reorientation, asymmetric body shapes, and internal propulsion~\cite{scholz2018inertial,loew:20,nosenko2020active,kim2021three,arkar2021dynamics}. The two types of inertial motion---translation and rotation---can lead to counteracting behaviors. Rotational inertia can enhance clustering and phase separation, whereas translational inertia tends to suppress aggregation~\cite{fily2017mechanical,lanoiselee2018statistical,caprini2021collective,te2021jerky,manacorda2017lattice,de2020phase,breoni2020active,mandal2019motility,scholz2018inertial,caprini2022role}. In cellular systems and dense active materials, inertia contributes to mechanical stress transmission, vortex-like flows, and contractile dynamics~\cite{taka:17,koch2021role}. At the macroscopic scale, actuated worms form entangled, motile structures~\cite{deblais2020rheology,deblais2023worm}, a behavior that has been successfully replicated using synthetically designed, soft-robot–based worm-blob systems~\cite{kaeser2025individual}. Such systems not only mimic entangled structures but also demonstrate large-scale collective locomotion~\cite{deblais2023worm}.

Furthermore, inertial effects play an important role in a broad class of synthetic active systems, ranging from self-propelled dust particles in a low-viscosity complex plasmas and granular particles on vibrating substrates to microflyers, automated flying drones, and their swarms~\cite{loew:20}. 
A promising polymer-type synthetic system is Kilobots, where soft-bodied robots are interconnected through elastic springs~\cite{pratissoli2019soft,baconnier2022selective,caprini2025spontaneous}. Such polymers, composed of chiral active beads, exhibit compression and spontaneous self-wrapping in two dimensions\cite{anand2024computer,caprini2025spontaneous}. For active Brownian polymer–like structures realized through linear assemblies of active droplets \cite{kuma:23}, paramagnetic microspheres \cite{garcia2018magnetically}, or active Janus colloids \cite{bisw:17}, inertial effects may be essential for their dynamics and structural behavior. }

Despite the seemingly fundamental role of inertia, its influence on the polymer conformations remains largely underexplored, particularly for long, closed-loop ring polymers. Ring-shaped polymers are prevalent in biological systems~\cite{trabi2002circular,dobay2003scaling,sanchez2010circularization}, including plasmids~\cite{lloyd2023microbial}, circular bacterial chromosomes~\cite{song:15,tubi:24}, contractile actin rings, circular RNAs~\cite{lasda2014circular}, and certain protein scaffolds~\cite{meaburn2007chromosome,deutsch1999equilibrium,lasda2014circular,greene2017circular}. Owing to their topology and absence of free ends, ring polymers exhibit distinct dynamical and structural features, such as self-threading~\cite{michieletto2014dynamics}, looping~\cite{obukhov2014melt}, a tank-treading  motion~\cite{phil:22,winkler2024active}, and the formation of topological glasses and topologically protected structures~\cite{muller1996topological,lang2012effect,halverson2011molecular}, characteristics that are inaccessible to their linear counterparts. Although in these examples, inertia may be of minor relevance, they highlight particular features of the ring geometry that will also be decisive for its conformational and dynamical properties in the presence of inertia.  

The goal of the present study is to elucidate the influence of translational and rotational inertia on the conformational and dynamical properties of an unknotted active Brownian ring polymer (ABRP+TRI). In particular, we aim to shed light on the mechanisms responsible for the stability of the contracted conformations of such polymers. For this purpose, a generic model of an active polymer is applied, with linearly connected active Brownian beads~\cite{ghosh2014dynamics,kaiser2015does,eise:16}. {\cblue The rotational motion of the latter is temporally correlated, both in the under- and overdamped case; therefore, their active contribution in the translational motion is determined by colored noise rather than white noise~\cite{gomp:16,eise:16}, and the fluctuating-dissipation theorem as well as detailed balance are violated~\cite{cate:12,cate:15}.} Systems of purely repulsive active Brownian particles exhibit a motility-induced phase transition (MIPS) already in the limit of an overdamped translational and rotational motion~\cite{redn:13,cate:12,bial:13,wyso:14,bechinger2016active,elgeti2015physics}. It emerges from a density-dependent slowdown of their translation motion and a positive feedback mechanism~\cite{cate:15}. In dense polymer systems, the noncrossability (self-avoidance) of polymers implies geometrical constraints---entanglements---with a severe impact on their dynamics~\cite{doi:86}. The emerging reptation dynamics is very slow and may not overcome topological constraints, as observed in chromatin~\cite{rosa:08,halv:14}.         

Our Langevin dynamics simulations demonstrate that the conformational and dynamical properties depend not only on the activity strength but also substantially on the bead mass and the moment of inertia. In particular, we find a non-monotonic structural response, with a re-entrant coil–globule transition with increasing activity. Notably, an increase in rotational inertia enhances polymer collapse, whereas greater translational inertia suppresses it and promotes swelling. We attribute the origin of the emerging tight globules to polymer self-avoidance~\cite{doi:86,flor:53} rather than MIPS, as the appearing polarization of the beads' propulsion directions points outward from the globule, in contrast to the dense phase in MIPS~\cite{fily:14.1,wyso:16}. Such a preference would resolve the dense phase of a MIPS cluster, but stabilize the globule in the presence of self-avoidance and entanglements. Only at larger activities are geometrical restrictions overcome. Moreover, the diffusive motion of a ring is significantly affected by rotational inertia. Specifically, the center-of-mass diffusivity increases with a power-law of the moment of inertia.

\begin{figure}[t]
{\includegraphics[width=\linewidth]{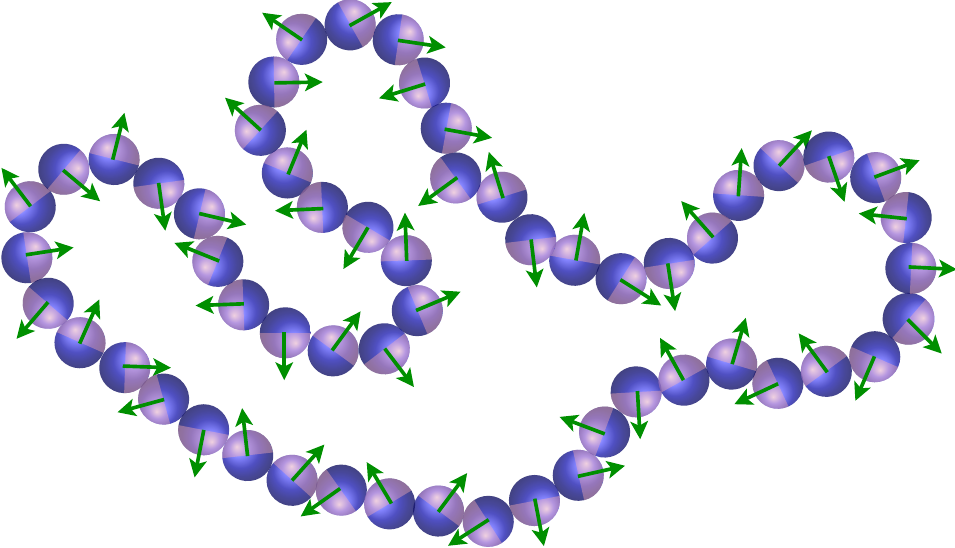}}
\caption{\textbf{Schematic.}
The schematic shows a model ring polymer comprised of inertial active Brownian particles (ABRP+TRI). The arrows indicate the propulsion direction, whose dynamics is governed by Eq.~\eqref{eq:eof_rotation}.  }
\label{Fig:schematic}
\end{figure}

\section{Methods}

The flexible active ring polymer is composed of $N$ beads at the positions $\bm r_i(t)$ with the velocities $\dot {\bm r}_i(t)$, $i \in  \{1,\ldots,N\}$, in three-dimensional space evolving with time $t$ (Fig.~\ref{Fig:schematic}). The equations of motion of the beads in the presence of translational inertia are given by the Langevin equations
\begin{equation} \label{eq:eof_translation}
    m \ddot{\bm r}_i = - \gamma \dot{\bm r}_i + \gamma v_0 \bm e_i + \bm F_i + \bm \varGamma_i ,
\end{equation}
where $m$ is the mass, $\gamma$ the translational friction coefficient, $v_0$ the constant active speed along the propulsion direction $\bm e_i$, $|\bm e_i|=1$, and $\bm F_i$ the total conservative force on bead $i$, which follows from respective potentials. The thermal environment is described by the forces $\bm \varGamma_i$, stationary Gaussian and Markovian stochastic processes with zero mean and the second moments
\begin{equation}
    \lla \varGamma_{\alpha i}(t) \varGamma_{\beta j}(t') \rra =  2 \gamma k_B T \delta_{\alpha \beta} \delta_{ij} \delta(t-t') 
\end{equation}
and $\alpha, \beta \in \{x,y,x\}$. Consecutive beads are connected by the harmonic bond potential 
\begin{equation} \label{eq:bond_pot}
    U_s = \frac{\kappa_s}{2} \sum_{i=1}^N \left(|\bm r_{i+1} - \bm r_i| -l_0) \right)^2
\end{equation}
of finite equilibrium bond length $l_0$ and the force constant $\kappa_s$. Due to the ring structure $\bm r_{N+1}= \bm r_1$. Self-avoidance is included by the repulsive, truncated, and shifted Lennard-Jones (LJ) potential
\begin{align} \label{eq:ev_pot}
\displaystyle
U_{LJ} = \left\{ 
\begin{array}{cc} \displaystyle
  4 \epsilon\sum_{i>j}^{N} \left[\left(\frac{\sigma}{r_{ij}}\right)^{12} -\left(\frac{\sigma}{r_{ij}}\right)^{6} + \frac{1}{4}\right], & r_{ij} \leq r_c  \\ \displaystyle
 0 , & r_{ij} > r_c 
\end{array}
\right. ,
\end{align}
where $r_c = \sqrt[6]{2} \sigma$ is the cut-off distance between the beads. Here, $\sigma$ is the LJ diameter of the beads, $\epsilon$ the interaction strength, and ${r}_{ij} = |\bm{r}_i-\bm{r}_j|$ is the distance between bead $i$ and $j$.

We describe the inertial dynamics for the propulsion direction in Cartesian coordinates via  
\begin{equation} \label{eq:eof_rotation}
    I \ddot{\bm e}_i = - \gamma_r \dot{\bm e}_i + \lambda_i \bm e_i +  \gamma_r (\bm \varGamma_i^e \times \bm e_i),  
\end{equation}
with the moment of inertia $I$, the rotational friction coefficient $\gamma_r$, and the active, non-thermal noise $\bm \varGamma_i^e$. The Lagrange multipliers $\lambda_i$ account for the holonomic constraints 
\begin{equation} \label{eq:const_e}
    \bm e_i(t)^2 = 1.
\end{equation}
Equations~\eqref{eq:eof_rotation} are generalizations of Lagrange's equations of the first kind~\cite{gold:80} by the present frictional and stochastic forces.
The formal elimination of the multiplier yields $\lambda_i = - I \dot{\bm e}_i^2$.  The active forces $\bm \varGamma_i^e$ are modeled as stationary Gaussian and Markovian stochastic processes with zero mean and the second moments
\begin{equation}
    \lla \varGamma_{\alpha i}^e(t) \varGamma_{\beta j}^e(t') \rra =  2 D_r \delta_{\alpha \beta} \delta_{ij} \delta(t-t') ,
\end{equation} 
where $D_r$ is the rotational diffusion coefficient  \cite{loew:20,sand:23}. The overdamped equations of motion follow easily for $I=0$, where, in particular,  $\lambda_i=0$.

The active rotational equations of motion, Eq.~\eqref{eq:eof_rotation}, violate the fluctuation-dissipation theorem, and the rotational diffusion and rotational friction coefficients are independent. Hence, the inertial persistent motion can be tuned independently by $I$ and $\gamma_r$.

The translational equations of motion are solved by the integration scheme proposed in Ref.~\cite{gronbech2013simple}. The detailed integration scheme for the equations of the propulsion directions is outlined in Ref.~\cite{tiwari2026virial} and in the SI, Ref.~\cite{supp} . \\

 {\it Parameters:} We present the simulation variables in reduced units. The lengths are expressed in units of the LJ parameter $\sigma$, energies in units of the thermal energy $k_B T$, and time in units of $\tau_r = 1/D_r$. The force constant of the bonds is chosen in the range $ \kappa_s = (5 \times 10^3 - 5 \times 10^4)k_B T/\sigma^2$ to prevent stretching of the bonds in the presence of large active forces. The bond length is set to $l_0/\sigma=1$ and the LJ energy to $\epsilon/k_BT =1$. The activity is presented by the dimensionless P\'eclet number $Pe= v_0/(\sigma D_r)$. The ratio of a beads translational, $D_0=k_BT/\gamma$, and rotational diffusion-coefficient is set to $D_0/(D_r \sigma^2)=1/3$. The translational and rotational inertia are incorporated through the reduced mass $M$ and reduced moment of inertia $J$, respectively, defined as
\begin{equation}
    M = \frac{m D_r}{\gamma} = \frac{\tau_m}{\tau_r} , \ \ \ J = \frac{I D_r}{\gamma_r} = \frac{\tau_I}{\tau_r}  ,
\end{equation}
with the characteristic times $\tau_m = m/\gamma$ and $\tau_I = I/\gamma_r$. 
The integration step is chosen in the range $\Delta t = (10^{-5} - 10^{-3})\tau_r$. Most of the simulation results are presented for the case of $ N=2048$ beads in the ring polymer. To determine scaling relations and scaling exponents as a function of the ring contour length, the bead number is varied from $N=256$ to $2048$.

The ratio of the dimensionless moment of inertia and mass,
\begin{equation}
    \frac{J}{M} = \frac{\gamma I}{\gamma_r m}  ,
\end{equation}
depends on the ratios of $I/m$ and $\gamma/\gamma_r$. As $\gamma$ and $\gamma_r$ may independently be adjusted, a large ratio $J/M$ can be achieved for a small $\gamma_r$, even if $I/m$ is small.

\begin{figure}[t]
{\includegraphics[width=0.275\columnwidth]{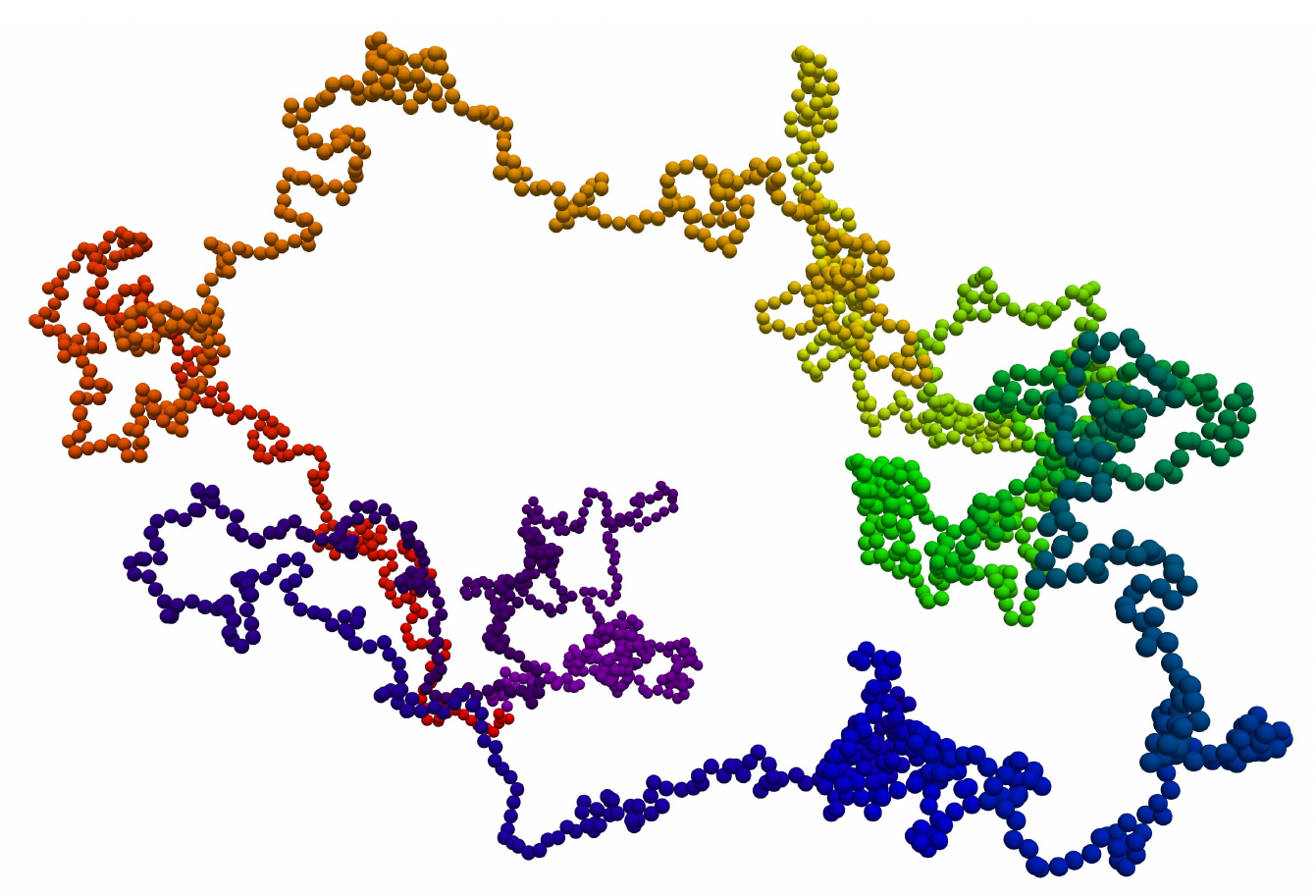}}
{\includegraphics[width=0.275\columnwidth]{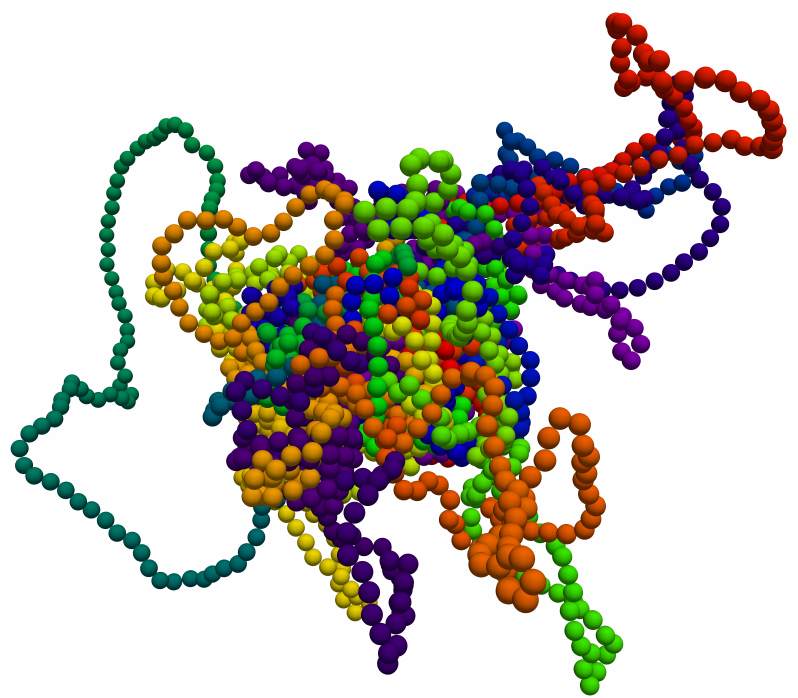}}
{\includegraphics[width=0.35\columnwidth]{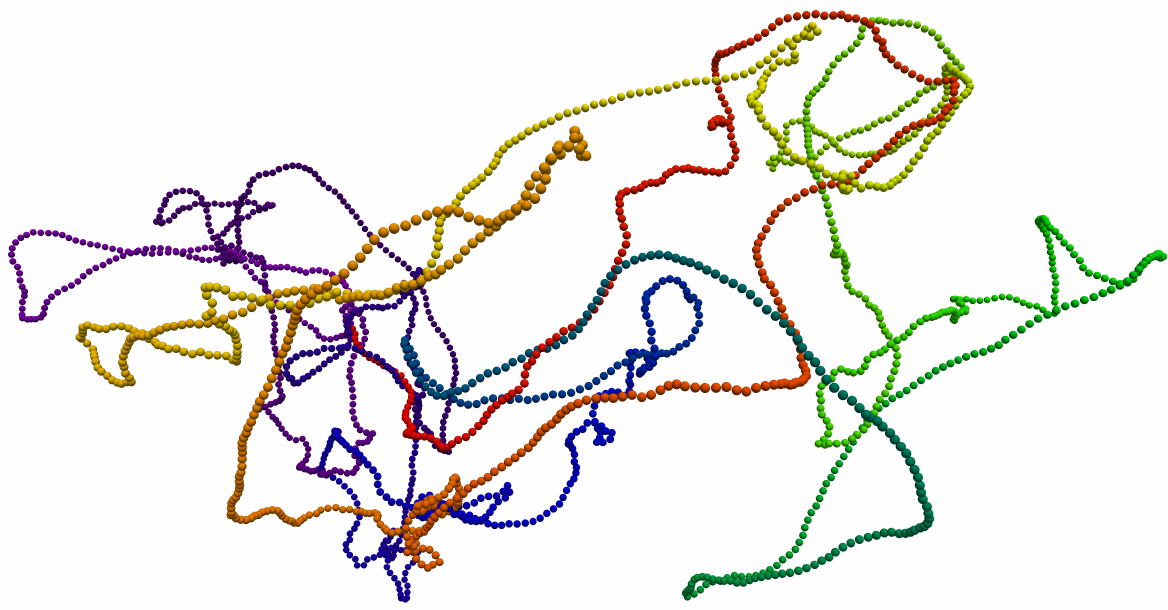}}
\caption{\textbf{Conformational state.}
Snapshots of different conformational states of a ring polymer for $Pe=1$ (left), $Pe=20$(center) and $J=100$ (right). The color gradient indicates the bead position along the contour. The other parameters are $M=0.3$, $J=5$, and $N=2048$ (cf. movies in Ref.~\cite{supp}).}
\label{Fig:conformation}
\end{figure}

\section{Results}

\begin{figure*}[t]
{\includegraphics[width=0.32\textwidth]{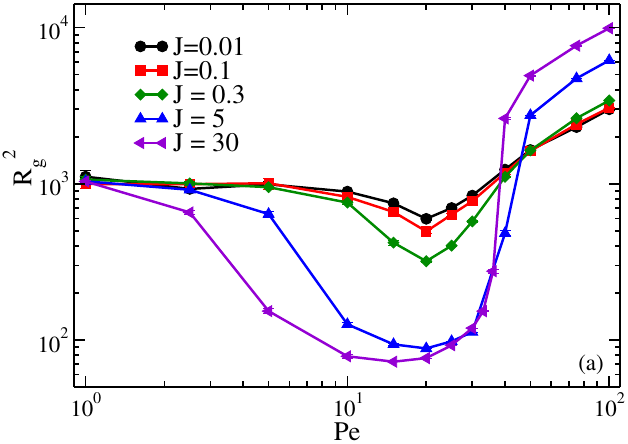}}
{\includegraphics[width=0.32\textwidth]{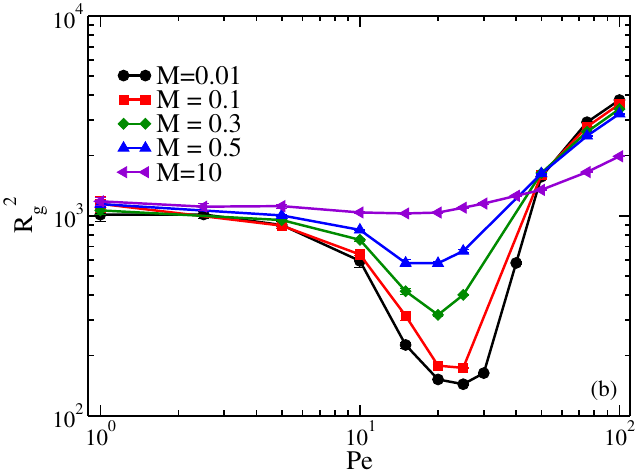}}
{\includegraphics[width=0.32\textwidth]{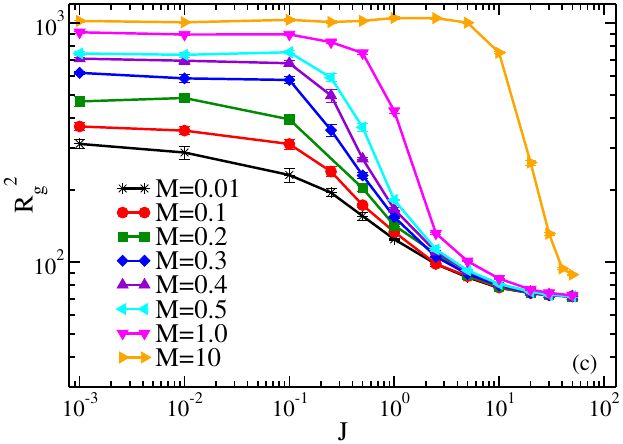}}
{\includegraphics[width=0.32\textwidth]{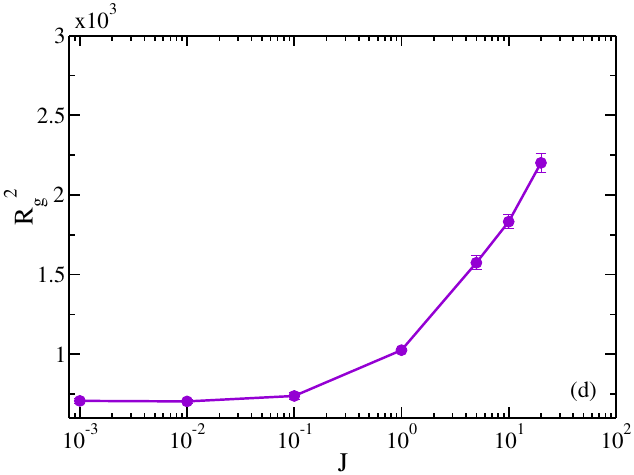}}
{\includegraphics[width=0.32\textwidth]{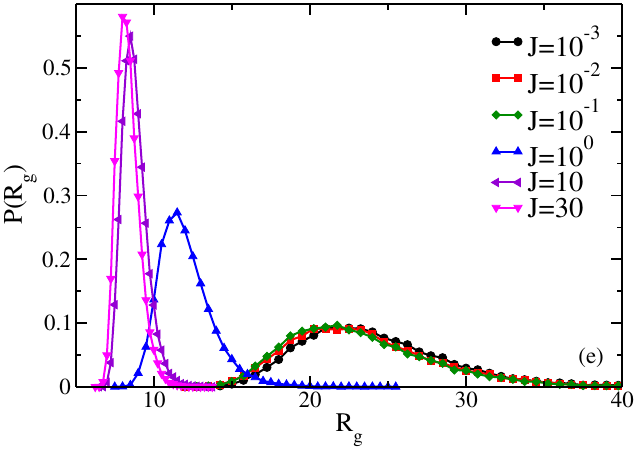}}
{\includegraphics[width=0.32\textwidth]{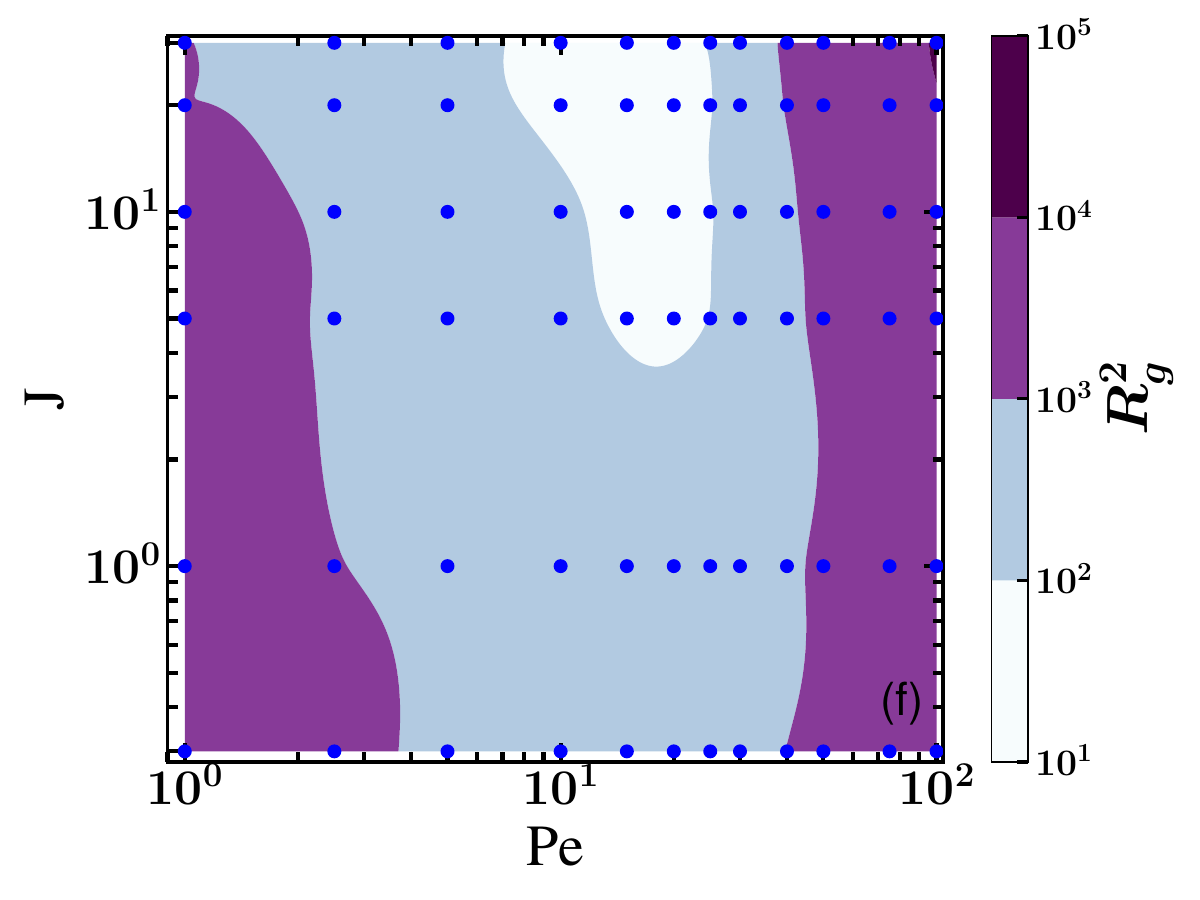}}
\caption{\textbf{Influence of inertia on active Brownian ring polymer.}
\textbf{(a)} and \textbf{(b)} Mean-square radius of gyration $R_g^2$ of an ABRP+TRI as a function of $Pe$ for various reduced moments of inertia $J$ and for various reduced masses at $M=0.3$ and $J=0.3$, respectively (cf. movie in Ref.~\cite{supp}). \textbf{(c)} Mean-square radius of gyration $R_g^2$ as a function of the reduced moment of inertia $J$ for various  reduced masses $M$.  \textbf{(d)} Mean-square radius of gyration of a phantom ring polymer as a function of $J$ at $M=1$. \textbf{(e)} Normalized probability distribution function of the radius of gyration for various $J$ and $M = 0.3$. In any case, $N=2048$, and in \textbf{d},\textbf{e}, and \textbf{f} $Pe=17$. \textbf{(f)} Phase diagram in the space of $J$ and $Pe$ at fixed $M=0.3$. 
 }
\label{Fig:radius_vs_pe}
\end{figure*}

\subsection{Radius of gyration: Coil-globule transition}

The conformational properties of the ABRP+TRIs{---\cblue illustrated in Fig.~\ref{Fig:conformation}---}are characterized by the mean-square radius of gyration $R_g^2$, defined as 
\begin{equation} 
    R_g^2 = \frac{1}{N}  \lla \sum_{i=1}^{N}  \Delta {\bm r}_i^2 \rra ,
    \label{Eq:R_g}
\end{equation}
with $\Delta \bm r_i ={\bm r}_i - {\bm r}_{cm}$ and the center-of-mass position ${\bm r}_{cm}$ of the ring. Figure~\ref{Fig:radius_vs_pe}(a) displays radii of gyration as a function of the P\'eclet number $Pe$ for various $J$ and $M$ values.
In the passive limit, $Pe \ll 1$, the ring polymer assumes a coiled state, which is predominantly governed by the excluded-volume and entropic interactions. As the activity increases, the polymer begins to shrink, approaching a minimum $R_g^2$ value at an intermediate activity. Upon further increase in activity, the ring swells again \cite{mous:19}. This non-monotonic behavior of the ring size becomes more prominent with increasing reduced moment of inertia $J$. Most importantly, at large $J \gtrsim 5$, the overall shrinkage is enhanced, and $R_g^2$ becomes more than an order of magnitude smaller than its equilibrium value, indicating globule-like conformations in this regime.  The ring swells at larger $Pe$ is enhanced for all values of $J$ in the considered range of $Pe$, as shown in Fig.\ref{Fig:conformation}. Strikingly, we observe a sharp transition in the ring conformations from the globular to a swollen coil-like structure at larger moments of inertia $J \gtrsim 30$ within a very narrow window of $Pe$ values. 

{\cblue Figure~\ref{Fig:radius_vs_pe}(b) displays the dependence of $R_g^2$ on $M$ for $J=0.3$.} The $R_g^2$ curves for the different $M$ values exhibit qualitatively a similar behavior to that observed for a variation in $J$ as a function of $Pe$. An initial shrinkage of the ring at intermediate activity is followed by swelling as $Pe$ increases. {\cmag In the absence of inertia, the shrinkage is caused by excluded-volume interactions,\cite{anan:20} which is also the predominant mechanism for the collapsed state for the small values $M=0.01$. Hence, the curve of $M=0.01$ is close to the overdamped dynamics of the ring. The extent of shrinkage decreases with increasing $M$, indicating that an increasing translational inertia suppresses activity-induced shrinkage and leads to re-swelling of the ring. Ultimately, a ring only swells with increasing $Pe$ in the limit $M \gg 1$.  
Hence, an increase in $M$ implies an opposite response to an increase in $J$.  Unlike the dependence on $J$, the swelling regime for $Pe>40$ depends weakly on $M$ for $M<1$. However, for large $M>1$, the size of the ring is smaller $Pe>40$ and exhibits a clear $M$ dependence. }

{\cblue Figure~\ref{Fig:radius_vs_pe}(c) displays the dependence of the radius of gyration on the inertial parameters $M$ and $J$ for $Pe=17$.} The polymer attains nearly maximal compaction for this $Pe$ at the various $J$ and $M$ (cf. Fig.~\ref{Fig:radius_vs_pe}(a),(b)). {\cmag For $J \ll 1$, the radius-of-gyration grows with increasing $M$ from the compact state to the size of a passive ring. As long as $J/M \lesssim 1$, the polymer conformations are nearly unperturbed, while a ring undergoes a monotonic shrinkage with increasing $J/M$.}  Noteworthy, in the limit of large $J$, the curves for the different values of $M$ converge, and the rings assume a universal globular, nearly densely-packed state.

The presence of excluded-volume effects and polymer self-avoidance, i.e., the noncrossability of the polymer backbone, plays a major role in the emergence of the globular state. {\cblue As illustrated in the Fig.~\ref{Fig:radius_vs_pe}(d)} for a phantom polymer, i.e., the self-avoidance interactions between the beads in  Eq.~\eqref{eq:ev_pot} are absent,  $R_g^2$ monotonically grows with increasing $J$. Hence, activity rather leads to polymer swelling than to shrinkage~\cite{eise:16,wink:20}. The presence of geometrical restrictions is primary for the emergent globular state.

{\cblue The conformational fluctuations of the ABRP+TRIs are illustrated in Fig.~\ref{Fig:radius_vs_pe}(e)}  via the probability distribution of the radius of gyration $R_g=\sqrt{R_g^2}$. At small values of $J$, the distribution is broad and centered around a large $R_g$ value, corresponding to the coil-like conformations under good-solvent conditions. As $J$ increases, the distribution becomes narrower and the maximum shifts toward smaller values of $R_g$. Both aspects indicate the transition to a densely packed, globular state. 
{\cblue A detailed phase diagram of the reentrant coil-globule transition is presented in Fig.~\ref{Fig:radius_vs_pe}(f) in the $J$ and $Pe$ plane using the radius of gyration. The lighter region corresponds to the dense-globular phase, while the dark region indicates the coil phase.}

In summary, for large reductions in moments of inertia, an ABRP+TRI is highly sensitive to activity, showing a pronounced conformational response. It assumes a tight globular state in an intermediate $Pe$ regime, and exhibits significant swelling at larger values of $Pe$, with a steep, phase-transition-like increase of the radius of gyration. An increase in translational inertia reduces shrinkage, suggesting that translational inertia dominates rotational inertia. Hence, translational and rotational inertia play distinct yet interrelated roles in determining the conformational properties of active ring polymers under varying activity strength. 

Rotational inertia dominates a ring's conformational properties for $J/M>1$, and the globular structure becomes essentially independent of the reduced mass $M$ in the limit $J/M \gg 1$. The characteristic time for the decay of the correlation function of the propulsion direction increases with increasing reduced rotational inertia for  $J \gg 1$~\cite{tiwari2026virial,sandoval2020pressure,scholz2018inertial,sand:23} (cf. Dynamics section). {\cmag As a consequence, the dynamics of an ABP+TRI particle becomes more persistent, i.e., the persistence length increases,\cite{tiwari2026virial,sandoval2020pressure} which leads to swelling of phantom rings and a stabilization of the globular state in the presence of polymer self-avoidance.}

The fundamental role of excluded-volume interactions and polymer self-avoidance is highlighted by an emergent polarization in the globular state due to bead propulsion. We characterize the inhomogeneous orientation of the propulsion directions by the radial polarization density
 
\begin{equation}
p_e(r) = \frac{1}{r^2} \lla \sum_{i=1}^{N}  \frac{\Delta \bm r_i}{|\Delta {\bm r}_i |}\cdot {\bm e}_i \, \delta( r-|\Delta {\bm r}_i|) \rra ,
\end{equation}
where $\Delta \bm r_i$ is the bead position with respect to the center of mass, and $r$ is the radial distance.

\begin{figure}[t]
{\includegraphics[width=0.475\columnwidth]{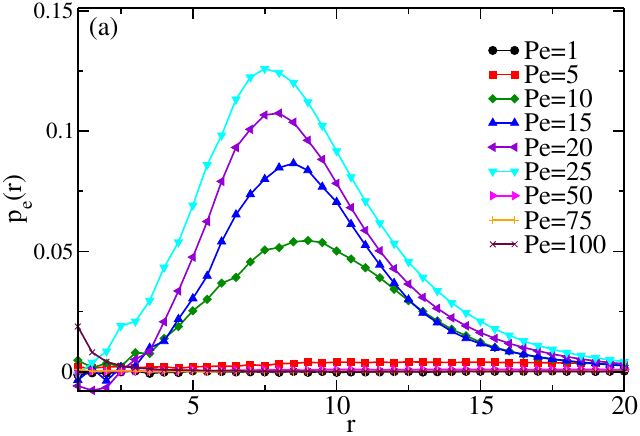}}
{\includegraphics[width=0.475\columnwidth]{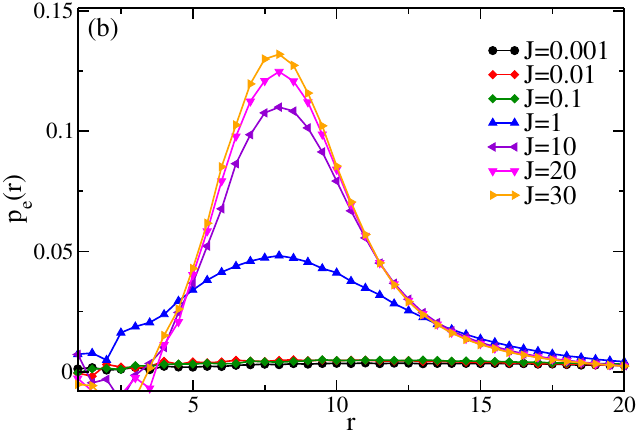}}
\caption{\textbf{Radial polarization density.}
Radial polarization density $p_e(r)$ with respect to the center of mass of a ring;  \textbf{(a)} for $M=0.3$,  $J=5$, and various $Pe$ (legend); \textbf{(b)} for $M=0.3$, $Pe=17$, and various reduced moments of inertia $J$ (legend) at $N=2048$.  }
\label{Fig:polarization_density_j5}
\end{figure}

Figure~\ref {Fig:polarization_density_j5} depicts the polarization density for various $Pe$ and $J$. 
For small $Pe<5$, a ring adopts a coil-like structure, and the propulsion directions are randomly oriented, which results in an approximately zero polarization density, Fig.~\ref {Fig:polarization_density_j5}(a). 
As $Pe$ and $J$ increase, a pronounced polarization appears, reflecting a preferential alignment of the propulsion direction radially outward with respect to the center of mass of the rings. The polarization grows with increasing radial distance, exhibits a peak at a radius larger than the peak of the radial bead-density distribution, and declines to zero for larger radii.

The polarization density is fundamentally different from that appearing in motility-induced phase separation (MIPS), where a non-zero local polarization spontaneously appears near the boundaries of the dense and dilute phases; however, it points toward the dense phase \cite{fily:12,cate:15,wyso:16,ausc:21,maha:23}. This polarization generates a normal pressure, which {\cblue stabilizes} a MIPS cluster. However, an outward polarization would destabilize a globule, as is reflected in the swelling of phantom polymers. Hence, the coil-globule transition of the ring polymer cannot be explained in analogy to MIPS. The presence of bonds and self-avoidance leads to entangled structures, which are stabilized by activity. The moment-of-inertia-enhanced rotation relaxation time of the propulsion direction leads to strong internal stresses and long-time stable structures. Moreover, rebounds, which are characteristic of passive particle interactions, are suppressed by activity~\cite{loew:20}.          

\begin{figure}[t]
{\includegraphics[width=0.475\columnwidth]{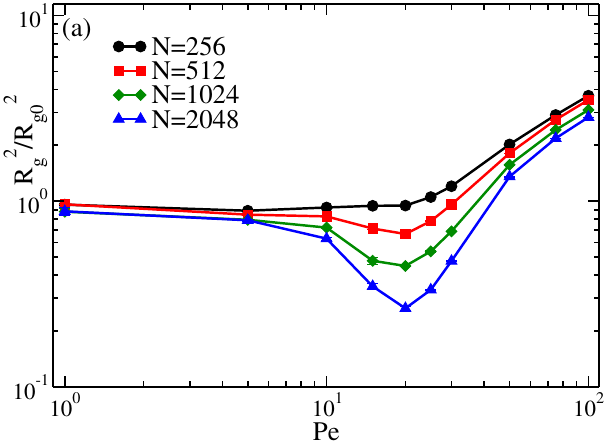}}
{\includegraphics[width=0.475\columnwidth]{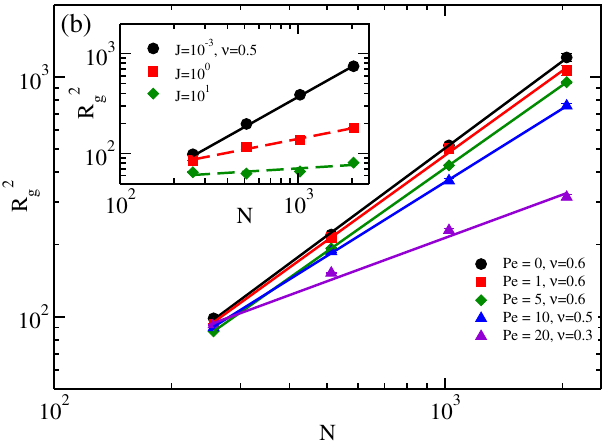}}
\caption{\textbf{Role of chain length and scaling exponent.}
\textbf{(a)} Mean-square  radius of gyration $R_g^2/R_{g0}^2$ as a function of $Pe$ for various numbers of beads, $M = 0.3$, and $J = 0.3$. Here, $R_{g0}^2$ is the radius of gyration of the passive ring polymer.  \textbf{(b)} Mean-square radius of gyration as a function of the number of beads $N$ for various $Pe$, $M = 0.3$, and $J = 0.3$. The inset in \textbf{(b)} shows $R_g^2$ for various $J$, $M=0.5$, and $Pe=17$.  The solid lines indicate the power-law dependence $R_g^2 \sim N^{2\nu}$.    }
\label{Fig:radius_n_vs_pe}
\end{figure}

\subsection{Radius of gyration: Scaling properties}

Polymer length-dependent physical properties can be derived using scaling relations~\cite{doi:86,dege:79}. Here, we address the activity dependence of the scaling of the radius of gyration with ring size. Figure~\ref{Fig:radius_n_vs_pe}(a) shows $R_g^2$
as a function of activity for various contour lengths.  For bead numbers $N \lesssim 300$ and the indicated $M$ and $J$ values, a ring swells monotonically with increasing $Pe$. An increasing ring size implies a non-monotonic behavior, where rings shrink in an intermediate $Pe$ regime. Notably, the shrinkage becomes more pronounced for larger bead numbers, as has also been observed for linear polymers in the overdamped limit, with and without bead attraction~\cite{anan:20,das:21}. The ring contour length has to exceed a certain value before geometrical restrictions substantially influence the polymer conformations. This is in line with our argument that self-avoidance plays a decisive role in the coil-globule transition, as studies on melts of linear polymers suggest that polymers have to exceed a minimal length before entanglements become relevant~\cite{doi:86,putz:00}.    

Figure~\ref{Fig:radius_n_vs_pe}(b) illustrates the scaling behavior of the active ring polymer with increasing contour length for various $Pe$, i.e, $R_g^2 \sim N^{2\nu}$, with the scaling exponent $\nu$. In the passive limit $(Pe=0)$, we find the well-known Flory exponent $\nu\approx 3/5$ of the three-dimensional self-avoiding random walk~\cite{doi:86}.  With increasing activity, $Pe=10$, the ring gradually shrinks, and the scaling exponent becomes $\nu \approx 1/2$, which corresponds to theta-solvent conditions and screening of the excluded-volume interactions~\cite{doi:86}. In the globular state at $Pe = 20$, we find the $\nu\approx 0.3$. Further increasing the $Pe$ leads to a re-swelling of the ring.  

The inset in Fig.~\ref{Fig:radius_n_vs_pe}(b) depicts the scaling properties for various $J$ values at $Pe=17$, i.e., close to the state with the highest compression. At lower values of $J=10^{-3}$, the scaling exponent $\nu \approx 1/2$ equals that of a phantom polymer~\cite{doi:86}.  However, as $J$ increases, $\nu$ decreases, and for $J=10$ the $R_g^2$ curve becomes nearly independent of $N$, which corresponds to a breakdown of scale invariance along the polymer contour. The weak dependence of $R_g^2$ on $N$ for larger $J$ reflects the fact that shorter chains do not exhibit significant conformational changes, while longer chains undergo substantial shrinkage. As the effective polymer size remains nearly unchanged, rings over a wide range of contour lengths can be packed within a comparable volume by increasing $J$.

\begin{figure}[ht]
{\includegraphics[width=0.475\columnwidth]{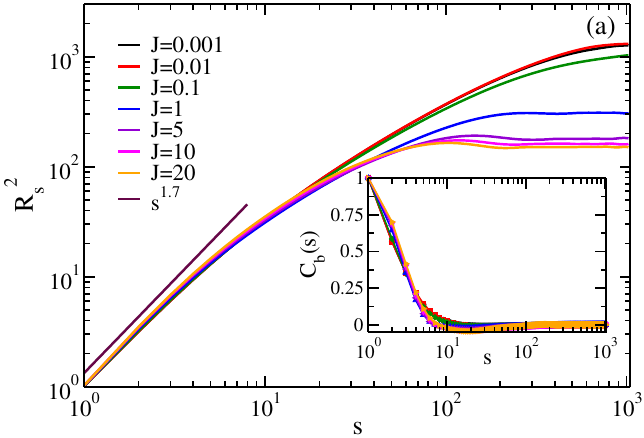}}
{\includegraphics[width=0.465\columnwidth]{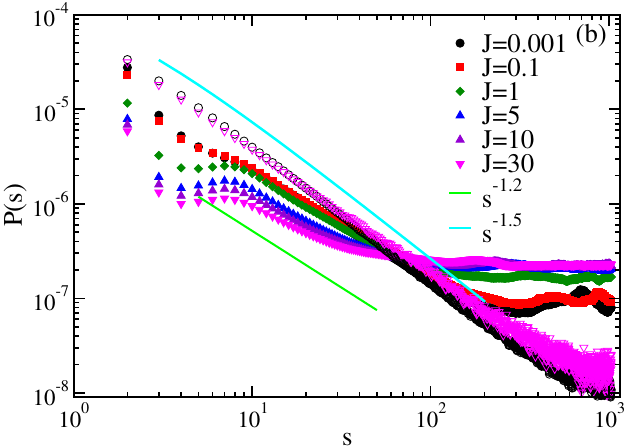}}
\caption{\textbf{Separation of beads along the contour.}
\textbf{(a)} Mean-square  distance $R_s^2=\langle ({\bm r}_i-{\bm r}_{i+s/l_0})^2\rangle$ between beads as a function of the their separation $s$ along the ring contour for various $J$. The inset displays the bond-vector correlation function $C_b(s) = \langle \bm R_i \cdot \bm R_{i+s/l_0}\rangle$ as a function of the contour distance.  \textbf{(b)} Normalized contact probability as a function of the contour length $s$ for various $J$ of self-avoiding (filled) and of phantom (open)  ABRP+TRIs. The other parameters are $M=0.2$, $Pe=17$, and $N=2048$.  }
\label{Fig:rij}
\end{figure}

\subsection{Internal Ring Structure}

\subsubsection{Contour Distance Probability}

Insight into the structure of the globular active ring polymer can be gained from the average geometric distance between two beads and their contact probability. Mean-square distances between beads,  $R_s^2=\langle ({\bm r}_i-{\bm r}_{i+s/l_0})^2\rangle$, separated by the contour distance $s$,
are presented in Fig.~\ref{Fig:rij}(a) for various values of $J$ and at $M=0.2$. The ring conformations exhibit no long-range self-similarity. 
 For contour lengths $s/l_0 < 20$,  $R_s^2$ is nearly independent of $J$, and for $R_s^2/l_0 <10$  it scales approximately as $R_s^2 \sim s^{1.7}$. Activity stretches the ring on this local scale, which is also reflected in the bond-vector correlation function $C_b(s) = \langle \bm R_i \cdot \bm R_{i+s/l_0}\rangle$ (inset of Fig.~\ref{Fig:rij}(a)). At small $J \lesssim 10^{-2}$, the mean-square distance grows approximately linearly with $s$ in the range $10< s < 2 \times 10^2$, corresponding to ideal (phantom) polymer behavior. Here, $C_b(s)$ is essentially zero.
Activity implies a screening of excluded-volume interactions and softens the local stiffer polymer on this length scale. For large length scales, the finite ring contour length is manifested, and  $R_s^2$ reaches a maximum.  As $J$ increase, particularly beyond $J\gtrsim 1$, a plateau appears for $s>10^2$. This saturation reflects the presence of a globular state. The polymer adopts a strongly folded conformation, so that the bead-bead distance no longer increases with the contour length. Particularly, distant beads along the polymer contour come into close proximity with roughly equal probability. The present folding is reflected in a negative bond-vector correlation function in the range $10 \lesssim s/l_0 \lesssim 10^2$  (inset of Fig.~\ref{Fig:rij}(a)).   

\subsubsection{Contact Probability}

To further characterize the ring's internal structure, we determine the contact probability between beads separated by the contour distance $s$. For this purpose, we assume that two beads are in contact if their spatial distance is within a cutoff distance of $r_c = 1.3\sigma$. Introducing the average number ${\cal N}(s)$ of beads within the cutoff separated by $s$ as   
\begin{equation}
    {\cal N}(s) = \frac{1}{N} \sum_{i=1}^N  \Theta\left(r_c - |{\bm r}_i - {\bm r}_{i+s/l_0}| \right),
\end{equation}
where $\Theta(x)$ is the Heaviside step function, the  contact probability distribution function is given by
\begin{equation}
    P(s) =  \langle {\cal N}(s) \rangle/{\sum_{s/l_0 =2}^N  \langle {\cal N}(s) \rangle}.
\end{equation} 
Here, the average corresponds to the ensemble average over various polymer conformations.

The contact probability shows a significant dependence on polymer self-avoidance and on rotational inertia, as is displayed in Fig.~\ref{Fig:rij}(b). For phantom active rings, which are slightly swollen for the indicated parameters, $P(s)$ decreases according to the power-law $P(s) \sim s^{-1.5}$ with increasing contour distance, independent of $J$. The exponent coincides with that of swollen chromatin in human embryonic stem cells~\cite{natesan2021active,lieberman2009comprehensive}. The local stretching of the self-avoiding ring polymer (Fig.~\ref{Fig:rij}(a)) leads to a stronger drop in the contact probability for small contour distances $s/l_0<10$. Compactification and folding for larger $J$ increases $P(s)$ again, and a maximum appears at $s/l_0 \approx 10$. In a $J$-dependent intermediate regime, $P(s)$ decreases with the approximate exponent $-1.2$. This exponent is larger than that of the phantom ring polymer ($-3/2$) and slightly smaller or even comparable with those of globular polymers~\cite{natesan2021active,lieberman2009comprehensive}. Finally, for large $s$, $P(s)$ assumes a plateau, which reflects the finite ring contour-length for $J \lesssim 10^{-2}$. For larger $J$, the plateau is consistent with the constant mean-square distance $R_s^2$, and again reflects the globular state of the ring polymer.

\begin{figure}[t]
{\includegraphics[width=0.5\columnwidth]{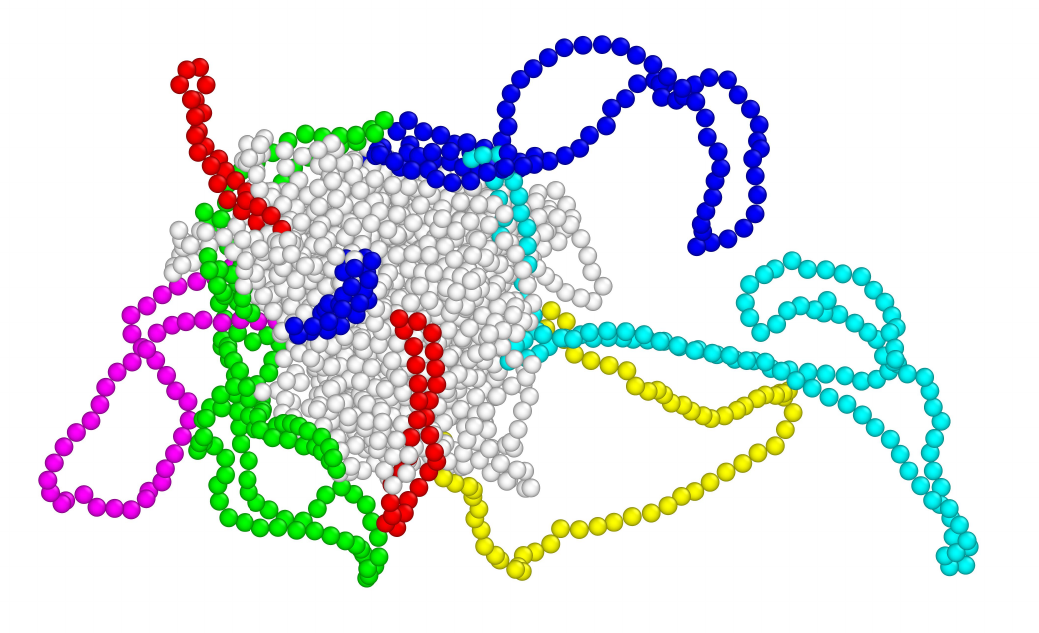}}
{\includegraphics[width=0.4\columnwidth]{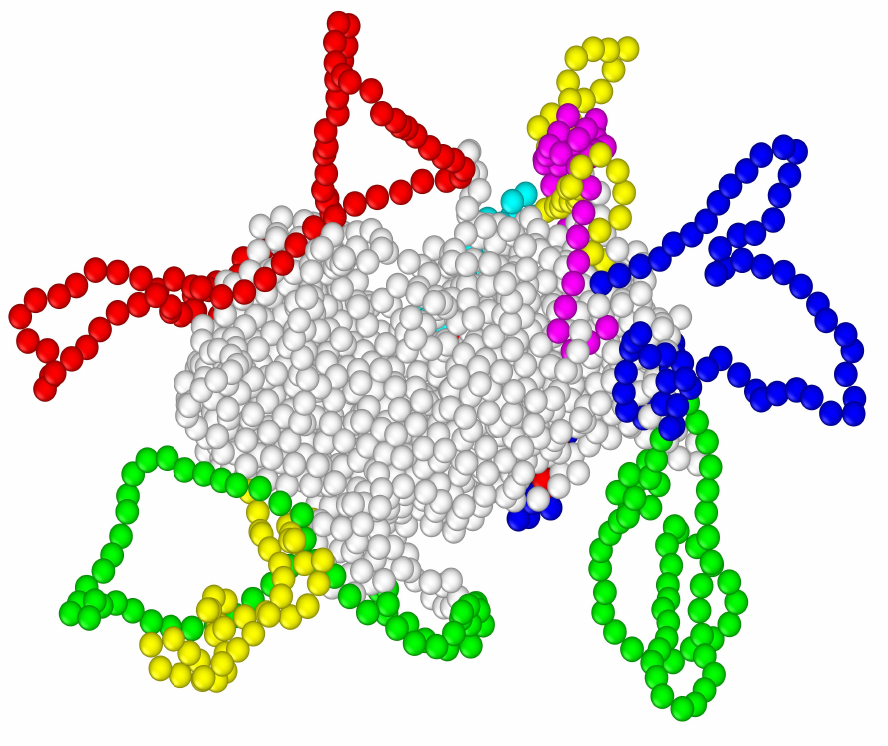}}
\caption{\textbf{ {\cmag Globular ring structures.}}
Snapshot of globular ring structures for $J=5$ (left) and $J=30$ (right). The globule part is indicated in gray, and the loops are shown in different colors. The other parameters are $M=0.3$, $Pe=17$, and $N=2048$. (cf. movie in Ref.~\cite{supp})}
\label{Fig:globule_loop_image}
\end{figure}

\begin{figure*}[t]
{\includegraphics[width=0.32\textwidth]{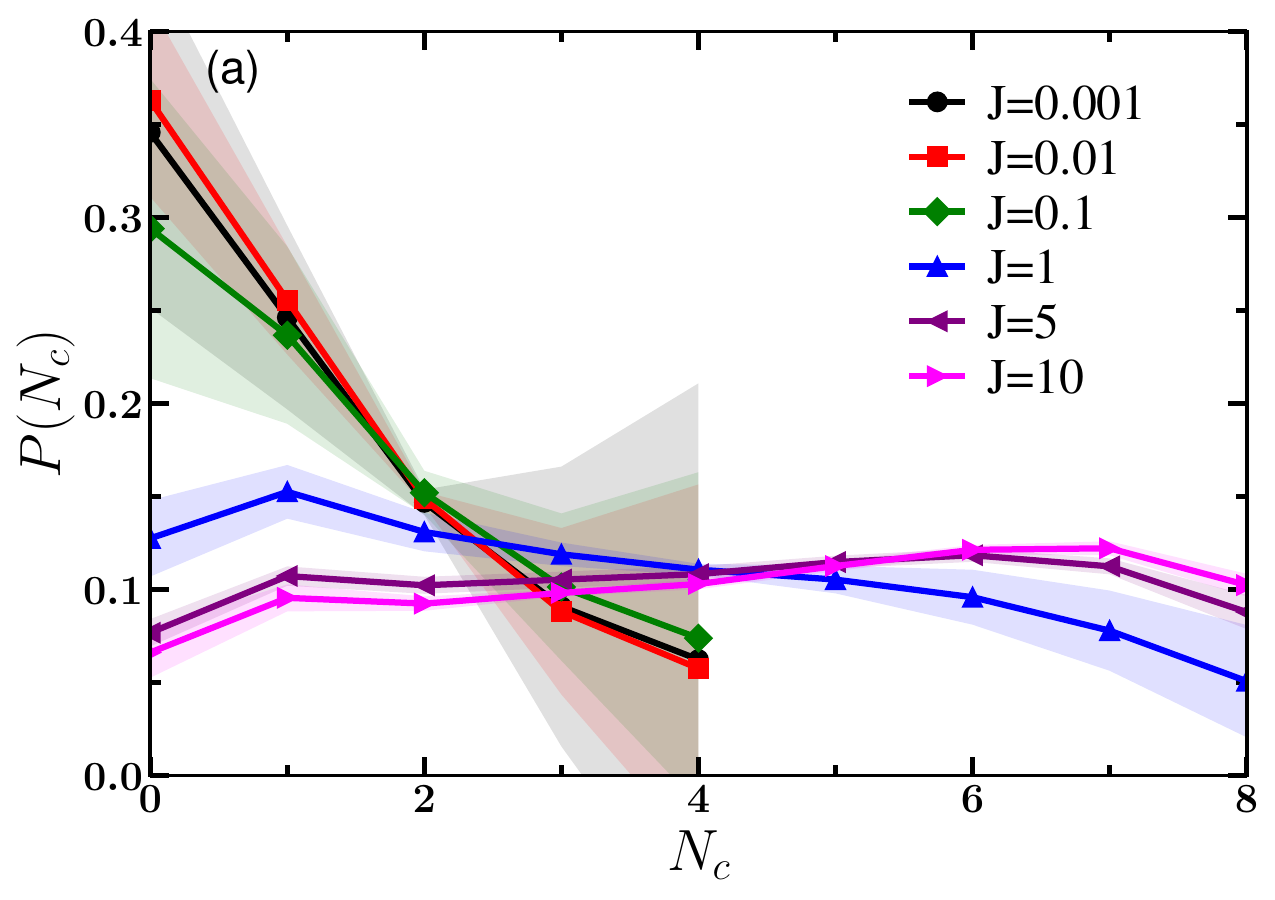}}
{\includegraphics[width=0.32\textwidth]{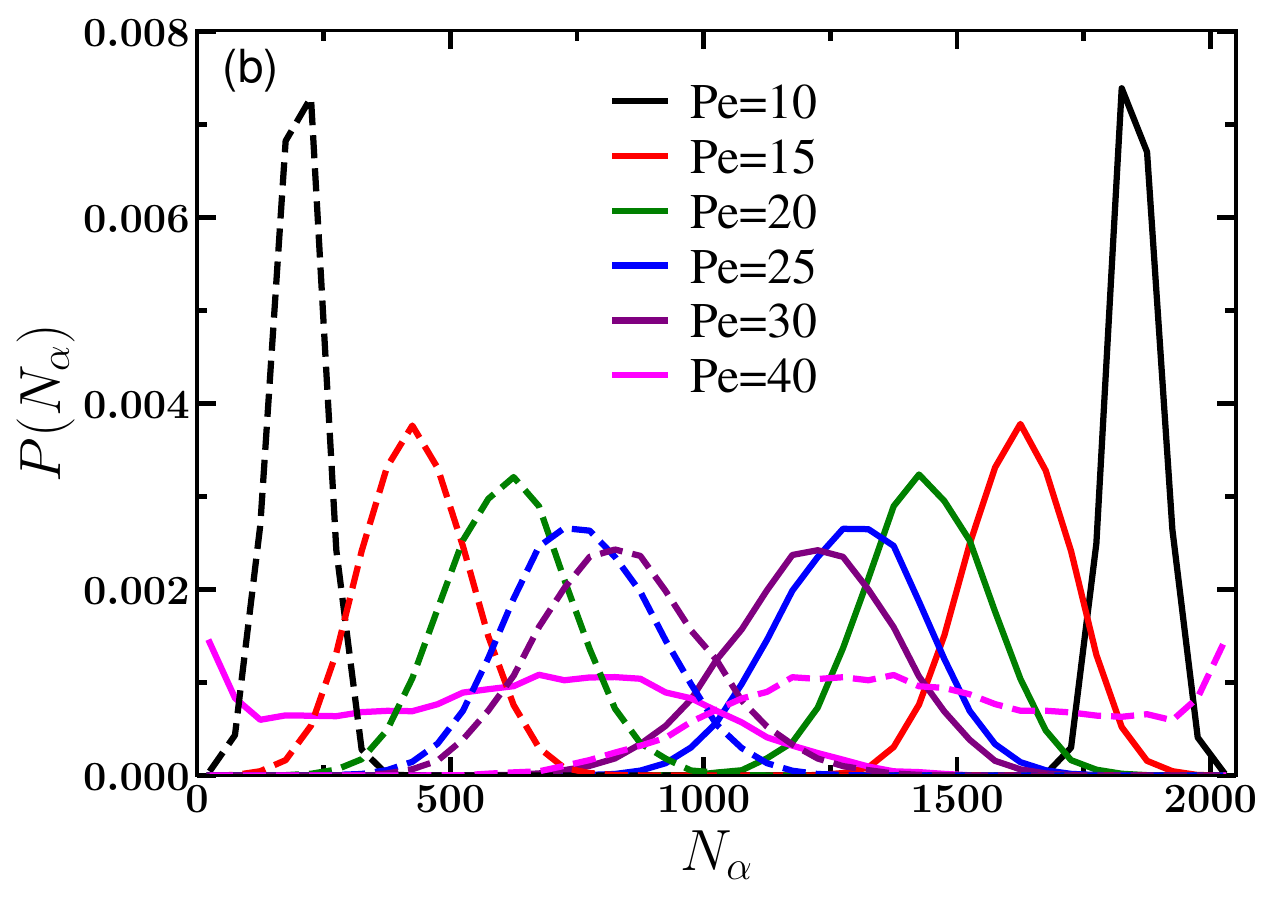}}
{\includegraphics[width=0.32\textwidth]{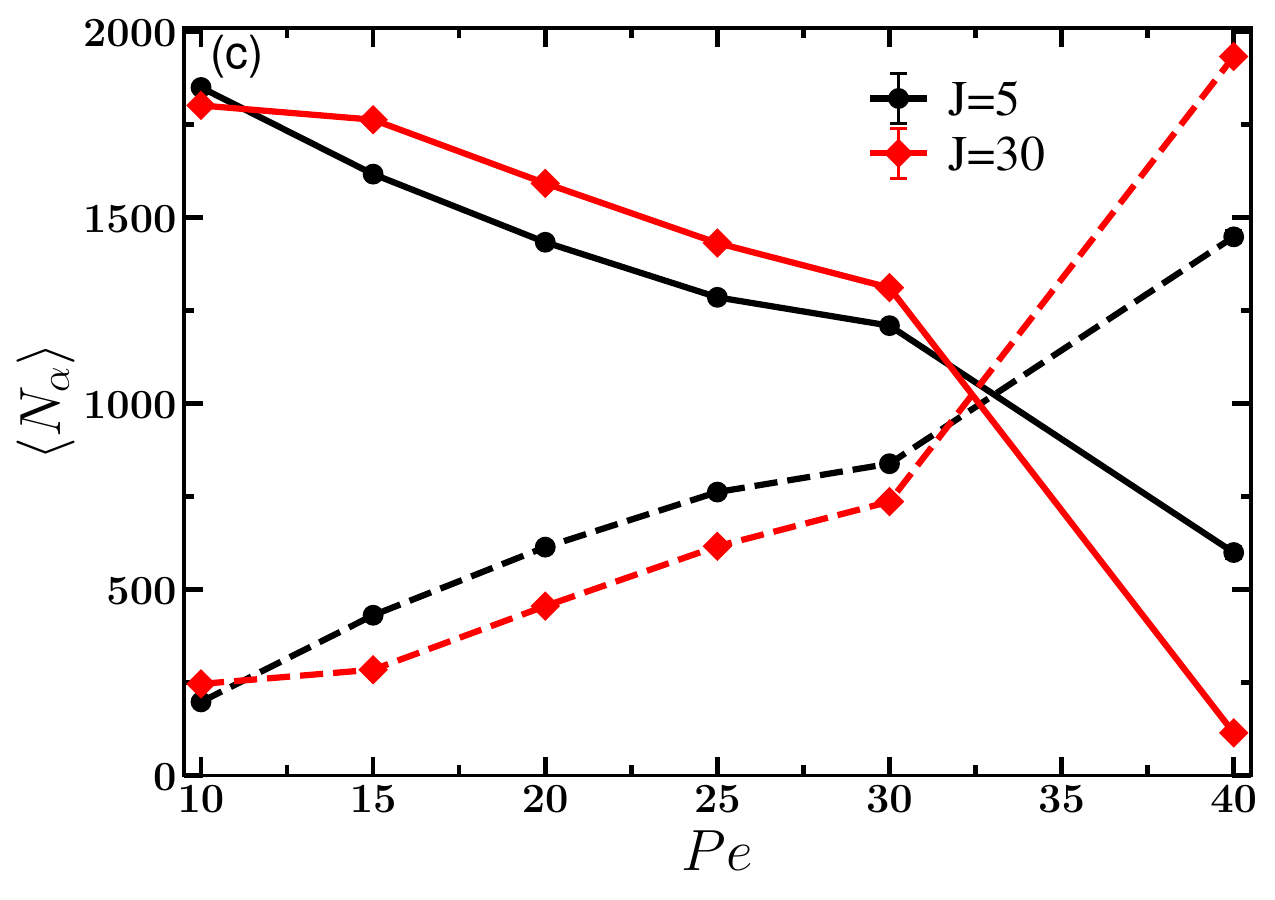}}
\caption{\textbf{Coordination number and distribution of beads.}
\textbf{(a)} Probability distribution of the coordination number $N_c$  for various $J$ at $Pe = 17$. The shaded areas indicate the statistical error of the distribution function.  \textbf{(b)} Normalized probability distribution of the number of beads $N_{\alpha}$ in the globule ($N_{g}$, solid) and the loops
($N_{l}$, dashed) for $J = 5$ at various $Pe$. \textbf{(c)} Average number of beads in the globule (solid) and the loops (dashed) as a function of the $Pe$ for $J = 5$ (black) and $J=30$ (red). The other parameters are $M=0.3$ and $N=2048$.
}
\label{Fig:coordination_m0.2}
\end{figure*}

\subsection{Globule Structure}

Figure~\ref{Fig:globule_loop_image} presents snapshots of globular ring conformations for two different values of the moment of inertia. Evidently, the presence of activity prevents the formation of a nearly homogeneous globule, but rather leads to a dense core and the emergence of loops of various lengths~\cite{buglakov2025motility}. 

\subsubsection{Coordination Number}

To characterize the globular structure, we determine the coordination number of the beads, independent of their location in a loop or the globular part. The probability distribution is defined as $P(N_c) = \langle N_c\rangle/\langle \sum_{N_c} N_c \rangle$, where $N_c$ is the average number of beads within the cutoff distance $r_c=1.3\sigma$ in the vicinity of every bead:
\begin{equation}
    N_c = \frac{1}{N (N-1)} \sum_{i=1}^N \sum_{j\neq i=1}^{N}  \Theta\left(r_c - |{\bm r}_i - {\bm r}_{j}| \right) ,
\end{equation}
and the brackets indicated the ensemble average over different realizations.
Figure~\ref{Fig:coordination_m0.2}(a) depicts the probability distribution $P(N_c)$ for various $J$ at $Pe=17$. For $J \lesssim 10^{-1}$, the distribution decays rapidly and vanishes for $N_c \gtrsim 5$, reflecting the loosely packed structure with a few neighboring beads in the coil-like ring conformation. In contrast, for $J > 1$, the distribution becomes nonmonotonic, very broad, with an approximately equal number of beads, and a slight peak at $N_c \approx 7$ before it drops for $N_c>8$. Hence, a bead in the coil part of the globule is frequently surrounded by many nearby neighbors. Yet, the structure is far from close-packed, as the coordination number of a 3D hexagonally close-packed structure is 12.  

\subsubsection{Loop Distribution}

As shown in Fig.~\ref{Fig:globule_loop_image}, the polymer forms multiple loops emanating from a single, globule-like core. To quantify these structures, we determine the number of beads in the globule-like phase and in the loops. A bead is classified as part of the globule if it has more than six neighbors within a cutoff distance of $1.5\sigma$.

Quantifying the loop size is more demanding because beads frequently attach and detach from the dense phase. Therefore, we impose a minimum loop-size criterion for loop identification. A connected sequence containing more than 24 beads is classified as a loop. They are highlighted in Fig.~\ref{Fig:globule_loop_image} in distinct colors.

Figure~\ref{Fig:coordination_m0.2}(b) displays the normalized probability distribution of the number of beads in the globule (solid lines) and in the loops (dashed lines) for $J=5$ at different activities $Pe$. Both distributions exhibit well-defined peaks for $Pe<40$, indicating that the majority of beads is confined in the globular phase. With increasing activity, the distribution of the globular phase shifts toward smaller beads, while the distribution of the loops shifts toward larger bead numbers, i.e., the globule resolves in favor of more and longer loops. 
The situation is reversed for $Pe=40$, where more beads are in loops than in the globular state. At larger $Pe$, the globule-coil transition occurs. 
This is supported by the $Pe$ dependence of the average number of beads in the globule, $N_g$ (solid), and in the loops, $N_l$ (dashed), shown in Fig.~\ref{Fig:coordination_m0.2}(c). As activity increases, the size of the globular phase decreases, while the average loop size increases, reflecting that high activity promotes polymer swelling through the formation of larger loops and a concomitant reduction in globule size. For $Pe \lesssim 35$, the qualitative behavior is the same for $J=5$ and $J=30$. However, at $Pe=40$ and for $J=30$, the global is essentially resolved, and almost all beads are in loops.

\begin{figure*}[t]
{\includegraphics[width=0.32\textwidth]{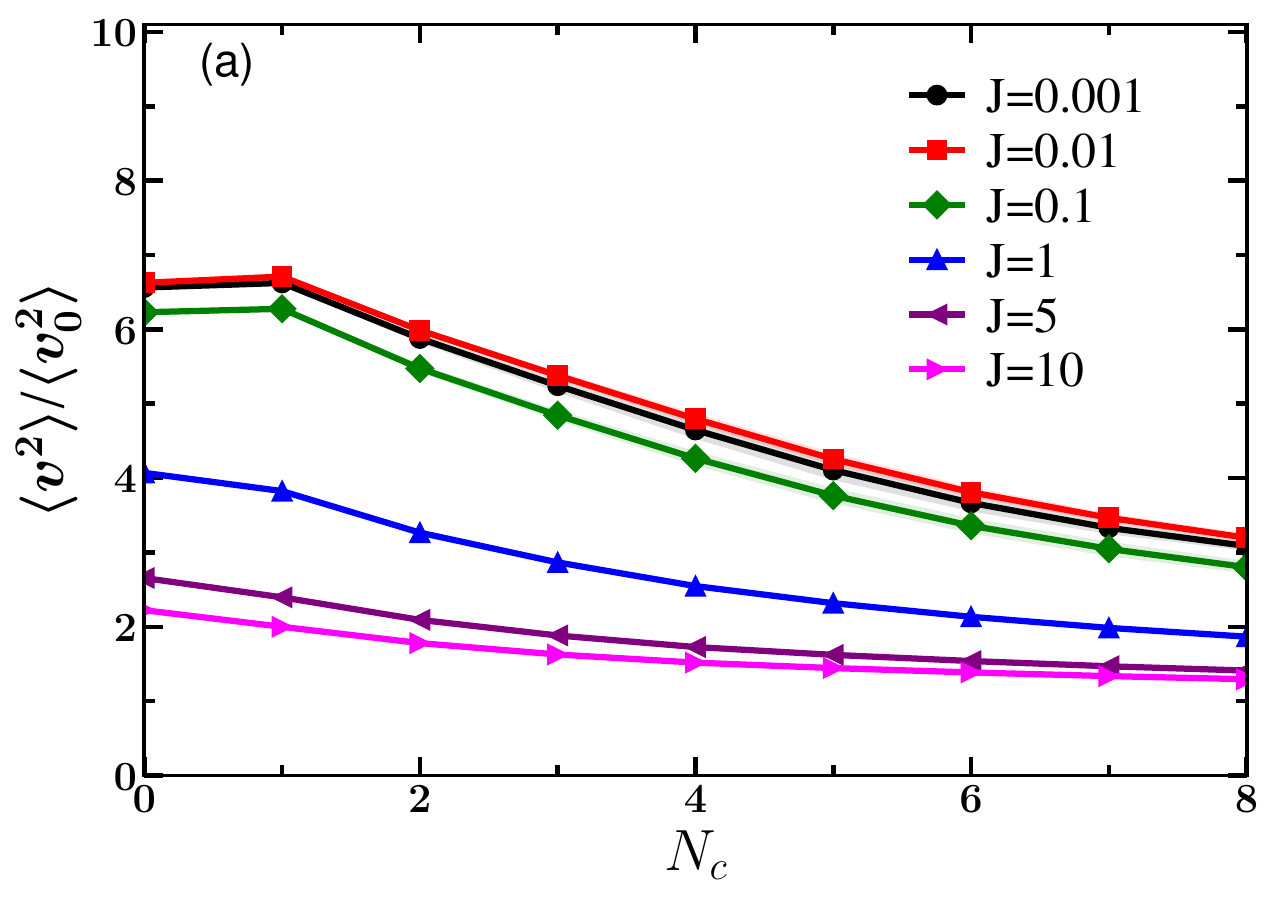}}
{\includegraphics[width=0.32\textwidth]{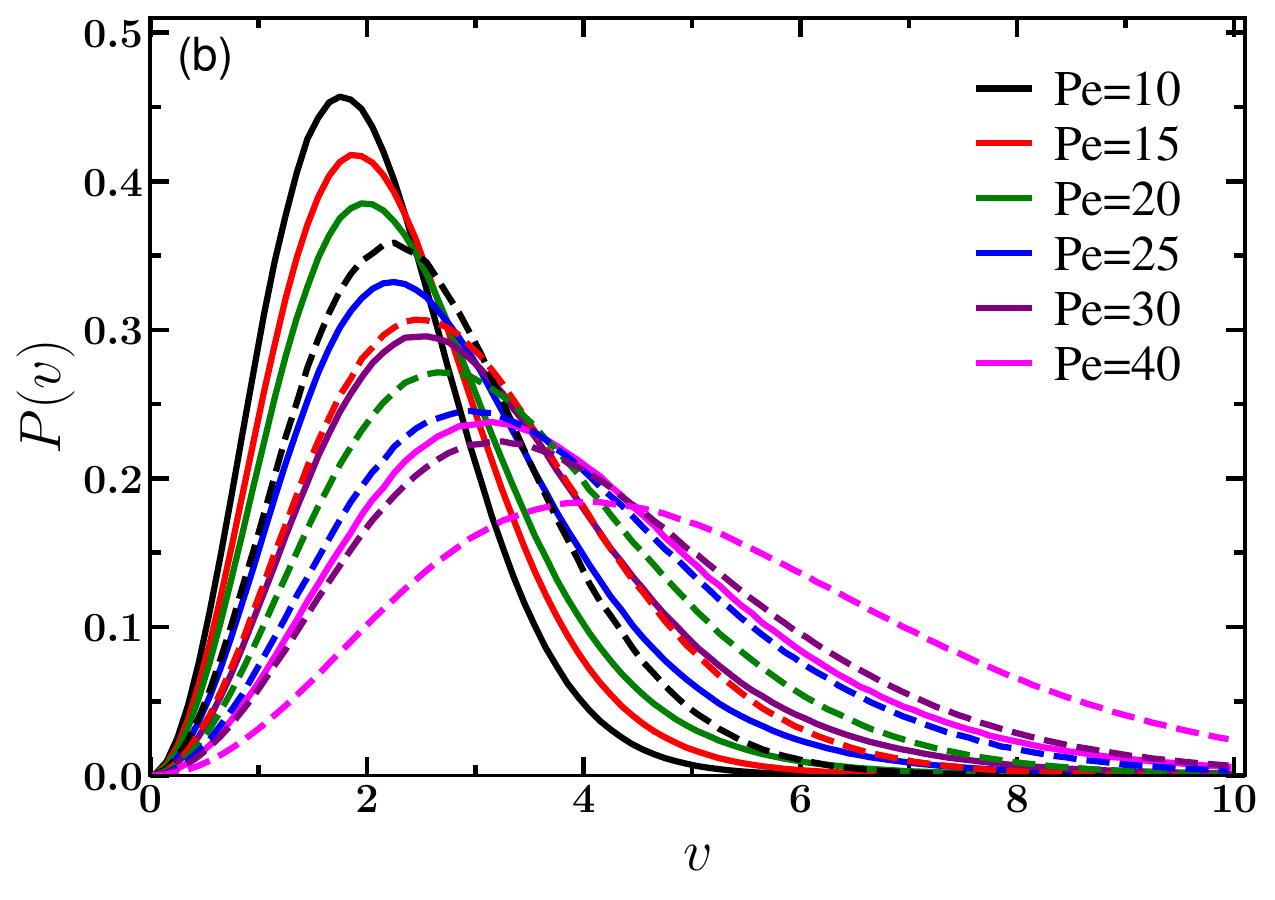}}
{\includegraphics[width=0.32\textwidth]{{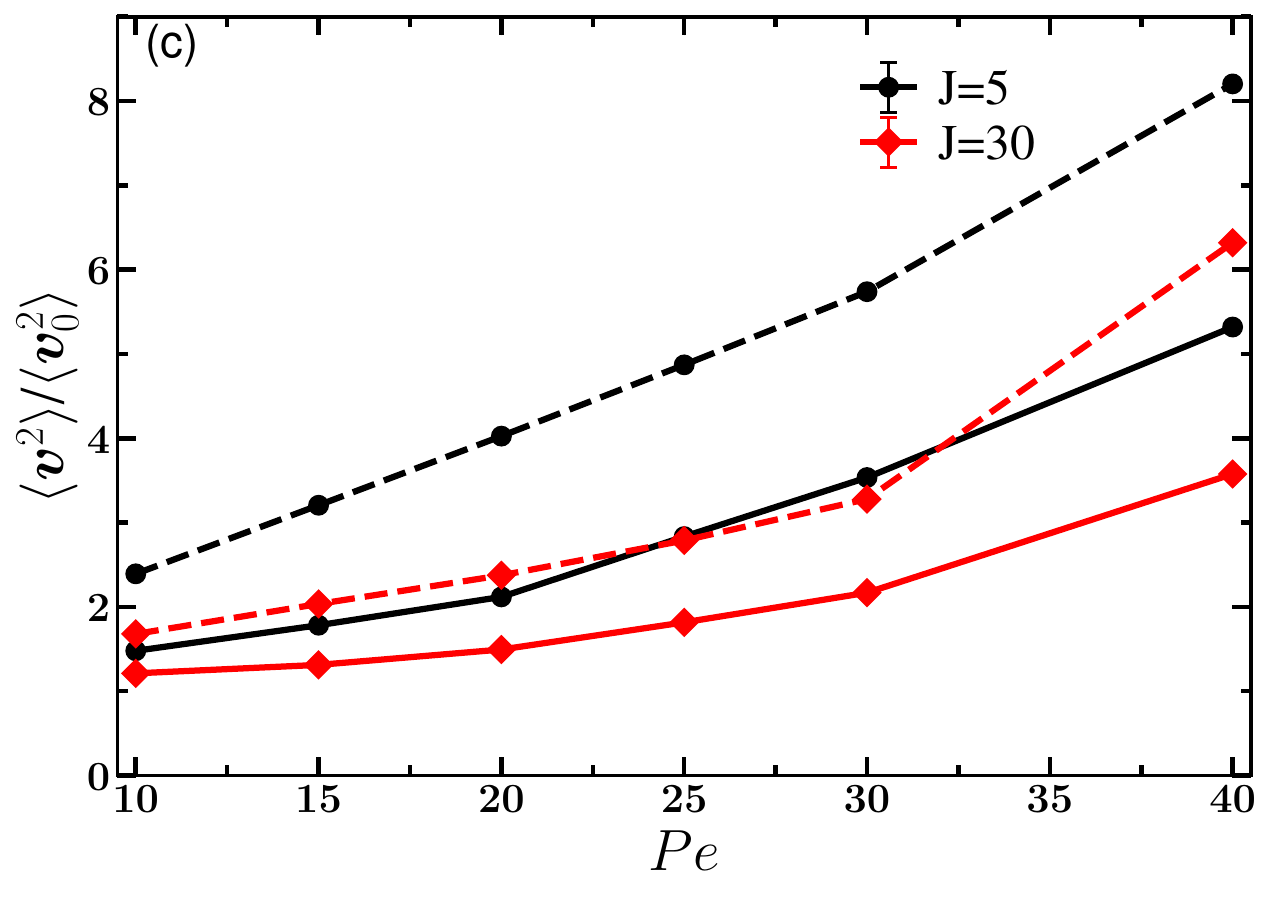}}}
\caption{\textbf{Average speed and distribution.}
\textbf{(a)} Mean-square bead velocity $\langle \bm v^2 \rangle$ as function of the coordination number for $Pe=17$ and various $J$. \textbf{(b)} Normalized speed distribution function for various $Pe$ and $J=5$. \textbf{(c)}  Mean-square bead velocity as a function of $Pe$ for various $J$. In \textbf{(b)} and \textbf{(c)}, the solid lines correspond to the globular and the dashed lines to the loop part of the globular ring. The mean-square velocity $\langle \bm v_0^2\rangle$ is the thermal velocity. The other parameters are $M=0.3$ and $N=2048$.}
\label{Fig:kinetic_energy}
\end{figure*}

\subsubsection{Kinetic Energy}

The key mechanism underlying MIPS is a feedback loop between the slowing of particles in high-density regions and the tendency of slowly moving particles to accumulate~\cite{cate:13}. Such a slowdown naturally appears in the globular state of the ring polymer, too. 

Figure~\ref{Fig:kinetic_energy} depicts the dependence of the speed of the beads on the coordination number,  as well as in the coil and loop part of the ring. At small $J<0.1$, the mean-square velocity is independent of the moment of inertia, and $\langle \bm v^2 \rangle$ decreases with increasing number of neighbors, which ``screen'' the active velocity~\cite{cate:13}. As the ring shrinks and the number of neighbors grows with increasing $J$, the active speed reduces and becomes nearly independent of the coordination number. {\cmag In fact, the actual speed is close to the thermal value for the large number of neighbors.} As expected, the normalized speed distribution of beads in the globule (solid) and in the emergent loops (dashed), presented in Fig.~\ref{Fig:kinetic_energy}(b), is shifted with respect to each other, with the peak in the globular phase at a lower speed than that of the loops. With increasing $Pe$, the peaks shift toward larger speeds, which indicates a gradual reduction of the geometrical restrictions. This is also reflected in Fig.~\ref{Fig:kinetic_energy}(c). Even more, this figure illustrates the effect of the moment of inertia on the mean-square velocity. At the smaller $J=5$, the speeds in the globule and loop phases are larger than for the larger $J=30$. The latter globule is denser, and the loops are shorter, resulting in stronger blocking of the bead motion and lower speeds. 

An increasing $Pe$ implies stronger active forces, which are ultimately able to overcome bead-bead blocking and the globule resolves. However, the mechanism of the coil-globule transition is primarily a consequence of polymer self-avoidance rather than the MIPS mechanism, as evidenced by the radial, outward-pointing polarization. Hence, the activity not only needs to overcome restrictions imposed by a dense neighborhood, but also those imposed by self-avoidance. Increasing activity leads to longer loops and higher bead speeds. As flexible phantom active Brownian polymers swell~\cite{eise:16,mous:19,wink:20}, swelling of the loops ultimately tears apart the globule, and the ring polymer swells with increasing $Pe$.  

\begin{figure*}[t]
{\includegraphics[width=0.32\textwidth]{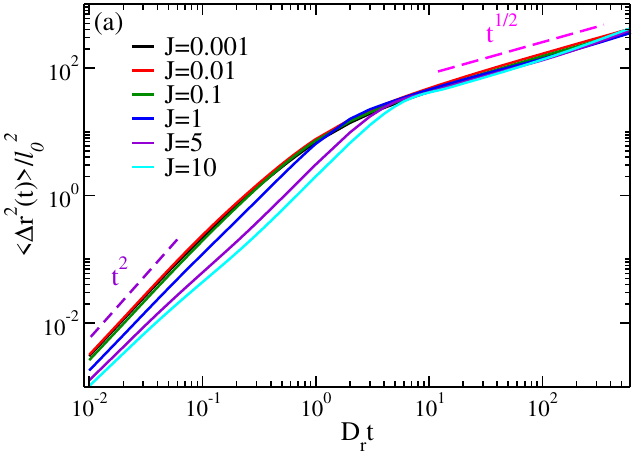}}
{\includegraphics[width=0.32\textwidth]{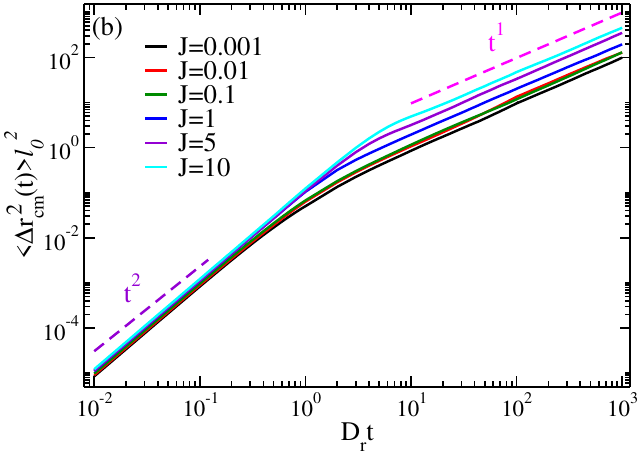}}
{\includegraphics[width=0.32\textwidth]{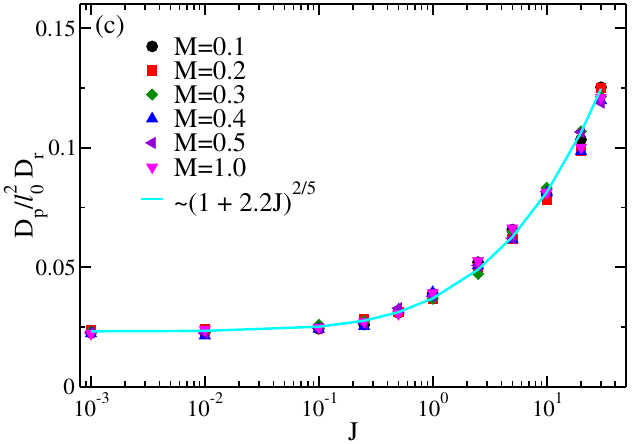}}
\caption{\textbf{Mean square displacement and diffusion.}
\textbf{(a)} Average bead and \textbf{(b)} center-of-mass mean-square displacement of the ABRP+TRIs for $M=0.1$ and various $J$. The dashed lines indicate power-laws for the respective time regimes. \textbf{(c)} Diffusion coefficients $D_p$  extracted from \textbf{b} as a function of the reduced moment of inertia $J$ for various $M$. The solid line indicates the dependence $D_p\sim (1+2.2J)^{2/5}$.   The other parameters are $Pe=17$ and $N=2048$.  }
\label{Fig:msd_total}
\end{figure*}

\subsection{Dynamics: Mean-Square Displacement} \label{sec:msd}

The dynamical features of the active ring polymer are quantified by the average mean-square displacement (MSD) of the beads  
\begin{equation}
    \lla \Delta r^2 \rra = \frac{1}{N} \sum_{i=1}^N \lla (\bm r_i(t) - \bm r_i(0) )^2 \rra
\end{equation} 
and the MSD of its center of mass (CM-MSD) $\langle \Delta r^2_{cm}(t) \rangle = \langle  ({\bm r}_{cm}(t)- {\bm r}_{cm}(0))^2 \rangle$, with the center-of-mass position $\bm r_{cm}(t)$.    

Figure \ref{Fig:msd_total}(a) displays the average MSD of the beads for $M=0.1$ and various $J$. Three time regimes can be identified: a moment-of-inertia-dependent ballistic regime, an indeterminate subdiffusive regime, and an asymptotic long-time diffusive regime by the total polymer (CM-MSD) for $D_r t > 10^3$ (not shown). The ballistic regime appears as a result of the active motion of the beads~\cite{eise:17,mous:19}, and the dependence on $J$ reflects the increasing confinement of the beads in the globular state. Remarkably, the intermediate time regime exhibits the time dependence of a phantom polymer, $t^{1/2}$~\cite{eise:17},  essentially independent of the moment of inertia. We attribute this effect to the strong active forces, which overcome excluded-volume restrictions and modify the modes of the flexible ring polymer~\cite{eise:17}.      

The center-of-mass dynamics of a ring is presented in Fig.~\ref{Fig:msd_total}(b). {\cmag The initial ballistic time regime is approximately given by ~\cite{tiwari2026virial}
\begin{equation}
\lla  \Delta r^2_{cm}(t) \rra \approx \frac{Pe^2 l_0^2}{N \left(1 + 2M/\sqrt{1+4J}\right)} (D_r t)^2 ,   
\end{equation} 
hence, depends only weakly on $J$ for the considered small mass $M$.
Yet, an increasing moment of inertia implies a shift of the crossover from the ballistic to the diffusion regime to longer times, as a consequence of the more persistent bead motion~\cite{caprini2022role,scholz2018inertial,sandoval2020pressure,tiwari2026virial}. } 

At long times $D_r t \gtrsim 10$, the bead and center-of-mass motion are diffusive.
We characterize the CM-MSD by the relation $\lla \Delta r^2_{cm}(t) \rra = 6D_p t$, with the diffusion coefficient
\begin{equation}
    D_p = D_p^0  + \frac{v_0^2}{6 N D_r} f(J), 
\end{equation}
where $D_p^0=k_B T/(N \gamma)$ is the center-of-mass diffusion coefficient of a passive polymer, and $f(J)$ a function of the moment of inertia. As displayed in Fig.~\ref{Fig:msd_total}(c), the diffusion coefficient is independent of the mass $m$, as proposed theoretically~\cite{sandoval2020pressure,scholz2018inertial}, and the simulation data are well fitted by $f(J) = (1+2.2J)^{2/5}$. 
Since the center-of-mass motion of the ring polymer is independent of the internal passive forces, the center-of-mass dynamics is equal to that of a free ABP+TRI, yet with the effective friction coefficient $\gamma N$. In case of an inertial ABP, theoretical considerations predict the asymptotic $J$ dependence $D_p \sim \sqrt{J}$ for $J \gg1$ in two dimensions~\cite{scholz2018inertial,loew:20,tiwari2026virial}. Our simulations predict a somewhat weaker increase. However, the qualitative trend is the same: the diffusion coefficient increases with increasing $J$ in a power-law manner.

\section{Conclusions}

In this article, we have systematically investigated the impact of translational and rotational inertia on various characteristic features of active Brownian ring polymers (ABRP+TRIs) using Langevin dynamics simulations. Our studies reveal an intriguing coil-globule-coil transition, driven by the interplay of activity ($Pe$), reduced mass ($M$), and reduced moment of inertia ($J$). The observed re-entrant transition is characterized by normalized distributions, the radius of gyration, contact probabilities, and inter-bead distance metrics.

We have demonstrated that the radius of gyration shows a non-monotonic dependence on activity, which becomes more pronounced as $J$ increases, emphasizing how rotational inertia promotes the shrinkage of the ring. The globular state can undergo re-expansion or re-swelling upon increasing activity. The analysis shows that while weakly active and passive rings exhibit a classical Flory-like scaling behavior of the radius of gyration with ring size $(\nu\approx 0.6)$, increasing activity and rotational inertia lead to phantom-polymer-like scaling $(\nu \approx 0.5)$, or the radius of gyration becomes even length independent,  which is indicative of a compressed state of the ring polymer.  
  
Rotational inertia implies a more persistent motion compared to overdamped ABPs, which generate stresses, suppresses configurational fluctuations, and promotes self-trapping. Even more, it leads, in combination with polymer entanglements, to an enhanced radially outward polarization of the propulsion directions {\cmag of the globular state} in the center-of-mass reference frame, which is responsible for the coil-globule transition. This contrasts with the motility-induced phase transition (MIPS) of ABPs, in which polarization points toward the dense region. In the ring coil-globule transition, topology is crucial, as self-avoidance and entanglements are fundamental. 

The globular state consists of two ``phases'', a dense core and loops, which form and disappear continuously. With increasing activity, the loop length grows, and with it the radially outward force---phantom active ABRPs swell monotonously with increasing activity~\cite{phil:22}. This leads to the disintegration of the globule and the emergence of a swollen ring configuration above a threshold activity level.   

The polymer's center-of-mass MSD exhibits a transition from a ballistic to a diffusive motion, which is typical of active matter systems with finite persistence. The translation diffusion coefficient is independent of translational inertia and follows approximately the power-law $D_p \sim J^{2/5}$ with increasing moment of inertia, confirming that rotational inertia plays a dominant role in regulating large-scale transport. The dependence of diffusivity on rotational inertia suggests a potential strategy for controlling polymer transport, while activity levels may saturate beyond a certain threshold, transport properties can still be effectively tuned via rotational persistence.

Our study provides key insights into how inertial effects---particularly rotational inertia---govern the properties of active Brownian ring polymers. The activity-induced compression of polymers has implications that extend well beyond a generic coarse-grained system considered here. Our findings contribute to a deeper understanding of synthetically designed ring polymers composed of granular rotors, or robotic active units, in which inertial forces play a crucial role. 

{\cblue 
Moreover, 
in living matter, the folding of the genome within the nucleus and the spatial segregation of active and inactive chromatin territories are believed to arise from nonequilibrium processes, including active forces, chemical gradients, and variations in effective temperature. In this context, the compressed conformations of active ring polymers---characterized by a dense core with emanating loop-like structures---provide a minimal physical framework that may offer new insights into the phase separation of active and passive polymer mixtures. 
More broadly, our results suggest that activity-driven compaction, mediated by connectivity and entanglements, can serve as a general mechanism for phase separation in polymeric suspensions. }

\section*{Data Availability}
The data that support the plots of this study are available from the authors upon request, while Supplementary Movie 1 and Supplementary Movie 2 are uploaded as Supplemental Material.

\section*{Code Availability}
The data are not publicly available. The data are available from the authors upon request.

\section*{Acknowledgment}
RP and SPS acknowledge financial support from the UGC-India and DST-SERB Grant No. CRG/2020/000661,  and computational time at IISER Bhopal and the  PARAM Smriti of NSM facility.

\section*{Competing Interests}
The authors declare no competing interests.

\section*{Authors Contribution}
RW and SPS designed and supervised the research project. RP and AP developed the simulation code, and RP has analyzed the simulation results. All authors contributed to the interpretation of the data, discussion of the results, and writing of the manuscript.

\bibliography{main,bibliography}

\end{document}